# Re.Dis.Cover Place with Generative AI: Exploring the Experience and Design of City Wandering with Image-to-Image AI

Peng-Kai Hung*

Industrial Design, Eindhoven University of Technology Eindhoven, The Netherlands; Department of Design, National Taiwan University of Science and Technology Taipei, Taiwan, p.hung@tue.nl

Janet Yi-Ching Huang

Industrial Design, Eindhoven University of Technology Eindhoven, The Netherlands, y.c.huang@tue.nl

Stephan Wensveen

Industrial Design, Eindhoven University of Technology Eindhoven, The Netherlands, s.a.g.wensveen@tue.nl

Rung-Huei Liang

Department of Design, National Taiwan University of Science and Technology Taipei, Taiwan, liang@mail.ntust.edu.tw

## Abstract

The HCI field has demonstrated a growing interest in leveraging emerging technologies to enrich urban experiences. However, insufficient studies investigate the experience and design space of AI image technology (AIGT) applications for playful urban interaction, despite its widespread adoption. To explore this gap, we conducted an exploratory study involving four participants who wandered and photographed within Eindhoven Centre and interacted with an image-to-image AI. Preliminary findings present their observations, the effect of their familiarity with places, and how AIGT becomes an explorer's tool or co-speculator. We then highlight AIGT's capability of supporting playfulness, reimaginations, and rediscoveries of places through defamiliarizing and familiarizing cityscapes. Additionally, we propose the metaphor AIGT as a 'tourist' to discuss its opportunities for engaging explorations and risks of stereotyping places. Collectively, our research provides initial empirical insights and design considerations, inspiring future HCI endeavors for creating urban play with generative AI.

## CCS CONCEPTS

Human-centered computing ~ Interaction design ~ Empirical studies in interaction design

---

* Corresponding author.

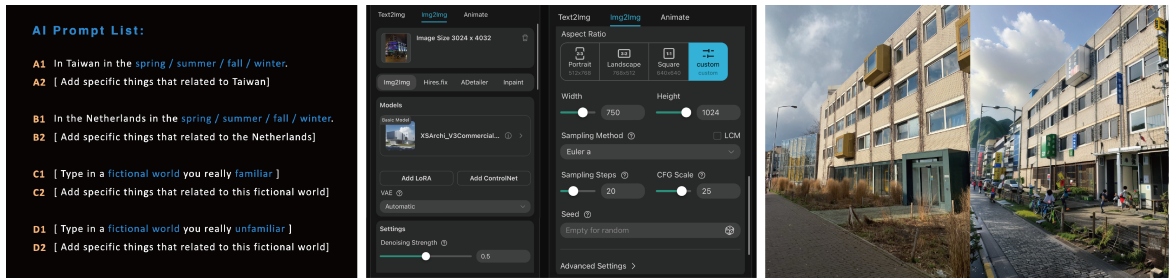

**Figure 1:** Left - The prompt list with four categories specified by the researchers. / Middle - The interface screenshots of Tensor Art website. / Right: An example of an input photo and an output AI image created by PT2. The image prompt: In Taiwan in the spring.

# 1 Introduction and related work

In the HCI field, the research space of Playable Cities and urban play have emerged and called for utilizing technologies to create rich urban experiences [1, 11, 12]. Instead of pursuing urban efficiency and economy, they view smart cities as a platform for various forms of play for city-making [1, 11], which allows urban space to benefit socio-cultural structures [12]. The forms of urban play vary, including tangible installations in public places, pervasive games, urban exploration practices, and other ludic technology-mediated urban interactions [1, 12]. They offer the possibilities of spontaneous or non-instrumental play [1], which allows citizens to obtain positive emotions and reframe tedious circumstances [12], encounter serendipitous discoveries [2], and reflect on urban lives [3]. While related research designs or commercial products employ technologies such as sensors and actuators in wearables and mobile devices, GPS, AR, and VR, considerably fewer studies focus on the playful experiences empowered by generative AI (GAI) in urban environment.

GAI refers to AI systems that produce new plausible content, such as texts, images, and sounds from existing data [6]. In contrast to classical machine learning merely classifying data to provide descriptions and judgment, GAI systems can generate creative and fresh results such as text, pictures, video, code, and diverse forms [6]. Recently, design research with GAI technologies has flourished in the HCI community [4]. Design researchers regard GAI tools as design materials for exploration and experimentation [4]. Inquiries into their diverse applications and roles, user experiences, and impacts on design process and society become a crucial consideration [6]. Within numerous GAI systems, AI image technology (AIGT) receives increasing attention due to its high accessibility and prevalence in people's daily lives [4]. It uses image synthesis models (e.g., Dall-E or Stable Diffusion) to generate images from images or text prompts [4]. Prior design projects used AIGT to create playful interactions that modify city scenes, such as Paragraphica [9], Onion AI [10], and Wander 2.0 [14]. In spite of the inspirational work they produced, the empirical results and analysis in their research are insufficient, necessitating further endeavors to better evaluate AIGT and offer comprehensive design principles.

In this work-in-progress, we aim to explore the experience and design space of applying AIGT to city exploration. This practice refers to wandering around urban areas without prior planning or optimized itineraries [13, 17]. When designing technologies for city exploration, familiarity is an essential factor [7, 16]. As a quality of everyday aesthetic experience, familiarity shapes the senses and meaning of places [8, 19]. Encountering a familiar or unfamiliar place could differ explorers' experiences, which informs design strategies [7, 17]. A series of studies have inquired into various technologies for enhancing physical or digital urban exploration, including mobile phones and place recommendation systems [5, 17], soundscape devices [7], virtual reality [16], and mixed reality [2]. The outcomes of these studies, along with the prominence of AIGT, inspire us to investigate the design potential of introducing AIGT to city exploration. Therefore, we probe into two research questions: (1) What are the experiences of exploring the city with AIGT? (2) What are the roles of AIGT in creating playful urban interaction? We undertook an exploratory study in which four participants with different familiarity of Eindhoven Centre wandered and photographed there and used image-to-image AI to generate images from their photos and specified prompts. Drawing from their creation and the subsequent interviews, we present how AIGT content facilitates place discoveries and rediscoveries, the effects of familiarity on the exploration process, and AIGT's roles as an exploration tool and co-speculator. Next, we discuss AIGT's capability of familiarizing and defamiliarizing places and its inherent unfamiliarity with input data, which implies design possibilities and concerns. Taken together, our study offers initial empirical contributions and design considerations that inspire future HCI work to enrich urban experiences with GAI.

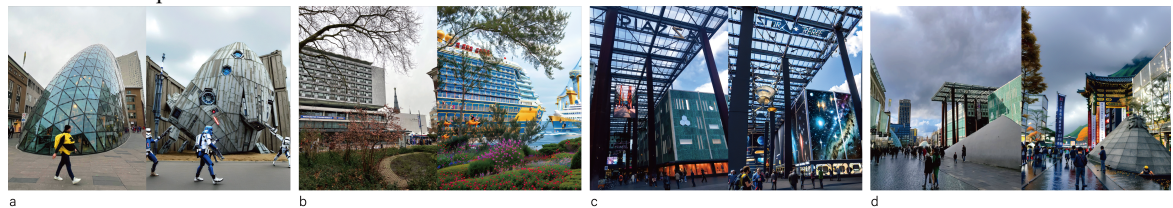

**Figure 2:** The participants' photos and AI images. The image prompts: a - Star Wars style. / b - Disney fantasy land. / c - Star Trek. / d - Taiwan in fall, during a festival.

## 2 Methods

In this exploratory study, two Taiwanese (PT1, PT2) and two Dutch (PD1, PD2) participants were recruited through social platforms. They age 25-31 years, with two males (PT1, PD1) and two females (PT2, PD2). Before joining the activity, they were briefed and signed informed consent approved by the Ethical Review Board. The longevity of their stay in Eindhoven ranges from 1 year (PT2, PD1), 3 years (PT1), and 10 years (PD2). The cultural backgrounds and longevity differ in their familiarity with places, which may affect the experiential outcomes of city wandering [7]. We expect that this difference could lead to a rich understanding of AIGT's experiences and its design space in urban exploration. The participants individually performed the exploratory activity for 2 hours. Firstly, they freely wandered the downtown region and took photos. When photographing, they were instructed to focus more on whole urban scenes instead of a single person or object. Then, they were asked to log in Tensor Art website [15] through the test account. This website features diffusion models that generate images from inputs such as images, text prompts, and parameters. During the exploration, the participants upload photos to the website, type in the prompts specified by the researchers, and experience the AI images.

Considering familiarity as a key quality in urban exploration [7] and AIGT's capability of integrating actual or fictional worlds, we planned the prompts based on these two factors to establish a basis for experience comparison (Figure 1, left). The prompts were categorized from A to D, indicating different degrees of familiarity with the worlds (e.g., A1 refers to an actual world familiar to PT1 and PT2 but relatively unfamiliar to PD1 and PD2). Each category consists of a prompt with a general description (A1-D1) and another one inviting them to fill in detailed descriptions (A2-D2). The participants were required to generate an image with each prompt. Additionally, they used a fixed GAI model and parameters (Figure 1, middle), which provided them convenience and avoided excessive uncertainty of generation. After the exploration, participants underwent semi-structured interviews which inquired into positive and negative parts of experiences, familiarity's effects, and reflections on their relationship with places and GAI. The interviews were audibly recorded, transcribed, and qualitatively analyzed using thematic analysis [18]. The first author reviewed all the photos, AI images, and interview data, identified potential themes and carefully discussed with two experts with backgrounds in computer science and design (i.e., the second and third authors), and established three themes presented in preliminary findings.

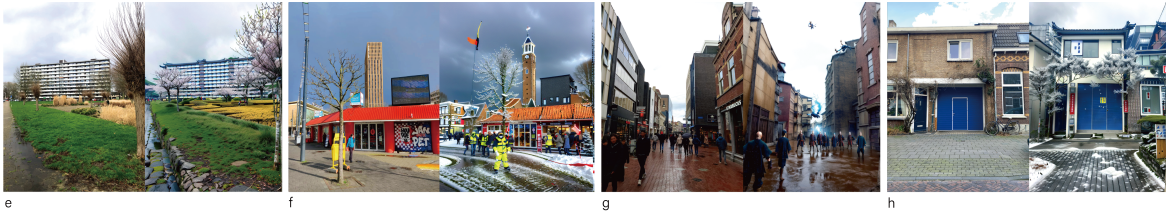

**Figure 3:** The participants' photos and AI images. The image prompts: e - In Taiwan in the spring with cherry blossoms. / f - The Netherlands in winter, during carnival. / g - Avatar the last airbender, Comet of Sozin is descending. / h - In Taiwan in the winter.

## 3 PRELIMINARY FINDINGS

### 3.1 Playfully Discovering and Rediscovering Place through AIGT Transformation

It was found that the participants were mainly engaged in (re)discovering place elements transformed by AIGT. AIGT's capacity to change existing buildings, passengers, and objects with fictional worlds themes sparked imagination, bringing novel discoveries of places. This characteristic was exemplified when PT1 took a landmark photo and generated a familiar fictional world rendition (Figure 2, a). PT1 praised AIGT's approach of "*keeping the building's shape while replacing its textures*" (PT1), enabling him to visualize it as Millennium Falcon and fantasize a Star Wars episode: "*It could be the Empire sending ships to colonize here... There are only three Stormtroopers now, but probably a whole row following them. It's their temporary base*" (PT1). He also emphasized the shifting tones and shades which altered the place ambiance (Figure 2, b): "*AI interpreted a gloomy, boring landscape into a vivid, lovely one*" (PT1). Interestingly, PT1 and PD2 expressed surprise at substitutions with entirely different functions, including a land into sea, a building into a ferry, and a store into a theater (Figure 2, c). As PT1 stated, "*it's impressive and satisfying because it's beyond my imagination*" (PT1).

When physical place appearances were integrated with other actual worlds, the participants tended to rediscover cultural and geographic elements, which brought playfulness, recollection, or facilitated reflection. PD1 noted the intriguing "*recontextualization*" (PD1) by AIGT which reshaped the shopping square into a festive bazaar (Figure 2, d). Although perceiving a sense of unfamiliarity, he felt "*really cool to see the building turned into a flag... and the big structure into ancient Asian-looking building*" (PD1). In PT2's case (Figure 1, right), the extruded structure modified by AIGT recalled her of "*something built illegally in Taiwan*" (PT2). Comparing the original architecture with its AI version made her ponder "*why a similar architectural approach was modernism in Holland, but AI interpreted it as a messy compromise in Taiwan*" (PT2). Moreover, PD2 pointed out the difference of architectural style when encountering a familiar place and its Taiwanese version (Figure 3, e). She critically conveyed her comment, drawing upon her knowledge of the location's historical background. "It's a special Asian roof… I especially like its details because the original building is quite ugly, very typical. It's one of the buildings from the '60s when they just needed much housing and chose easiest shapes" (P3). Though criticizing the place, PD2 rediscovered "how green it is" after noticing AIGT replaced its grass with unfamiliar plants. She concluded that AIGT enabled her to "view the familiar place through various lens and appreciate it more."

### 3.2 Familiarity as a Design Material for Generating AI Images and Creating Imaginary Worlds

The participants varied in their familiarity with where they explored and AIGT content, showing differences in their **(1) photographing and prompting strategies, (2) sensitivity to AI images, and (3) richness of imagination**. Firstly, the participants utilized their knowing of places to playfully capture photos and plan prompts. Inspired by the familiarity with local cultural events, PD1 creatively combined winter and carnivals in the prompt, indicating two elements that typically do not coincide in the Netherlands. PT2 explored landscapes blending Taiwanese and Dutch styles based on her memories of these countries, curious about "*how AI reinterprets them to be more Taiwanese or Dutch*" (PT2). PD2 chose to generate an image in a personally relevant location that she considered "*typical Dutch and had no element looked Taiwanese*" (PD2). She anticipated that AIGT would transform Dutch elements and beautify the scenery (Figure 3, e).

Regarding sensitivity to AI images, if AIGT integrated participants' familiar places, they provided detailed observations and gained insights into AIGT's perspective. PD1 easily identified classic and modern Dutch elements in an AI image (Figure 3, f). "*The original building is pretty industrial, but it turns into a tiled roof and clock tower, very classic old Dutch*" (PD1). He reflected that "AI's interpretation of the Netherlands is old Netherlands rather industrial ones" (PD1). Yet, he expressed disappointment upon receiving an image unrelated to his prompt (Figure 3, g): "*I described Avatar's final fight, but it didn't happen… It shattered my expectations negatively. I mostly expect it to churn out Avatar's animation style, but the model is trained on architecture and isn't familiar with Avatar's context*" (PD1). In parallel, based on the living experiences in Taiwan, PT1 and PT2 revealed that AIGT mistakenly depicted Taiwan's scene, prompting their reflections on AIGT's problems (Figure 3, h): "*AI puts snow in the place, but actually Taiwan's winter doesn't look like this. AI doesn't know the logic of the whole prompt sentence but combines the stereotype of winter with the stereotype of Taiwan*" (PT2).

Lastly, when AI images integrate the elements significantly familiar to the participants, they playfully expanded their imagination to the nearby region. The Star Wars episode fantasized by PT1 (Figure 2, a) and PD2's experience (Figure 3, e) show some illustrative examples. The fusion of elements in PD2's personal relevant place and those from Taiwan ignited her imagination beyond the AI image's border. "If there's a mountain, what would it look like? There's a church on this roadside, what would that appear in a Taiwanese style" (PD2). The outcomes of AIGT prompted her to "imagine what other stories could be from a different cultural perspective" (PD2).

## 3.3   Two Roles of AIGT in Urban Exploration: An Explorer's Tool and Co-speculator

The participants experienced two roles of AIGT in the study, implying the opportunities and challenges of incorporating it into urban play. Three participants regarded AIGT as **an explorer's tool** that infused familiar places with freshness and fun (PT1, PD1) or elicits reflections (PT2). AIGT playfully enlivens the mundane by entangling them with fictional worlds. Referring to Figure 2, PT1 stated that "*without AI, I wouldn't have thought this building like that. The real world is so far different from Star Wars that there's no trigger point, but AI helps to do that*" (PT1). Despite reduced emotional connections to familiar places, AIGT "*provides a new motivation to explore them*" (PT1). Additionally, PT2 underscored AIGT's role in facilitating reflections on places. "AI let me relate the places where I've lived and reflect more on what it means very Taiwanese or Dutch" (PT2). She emphasized the importance of AI prompts in this reflective process: "*It's because the prompt is Taiwan that I consider how the mountains in the image relate to this word… realizing that in Taiwan you can always see mountains if looking into the distance. It's quite different from the Netherlands*" (PT2). From her perspective, "*it's not just AI images that stimulate my thinking, AI prompts also lead me to look at images' results and reflect*" (PT2).

In contrast, PD2 viewed AIGT as "**a co-speculator** or creative buddy" (PD2). Due to AIGT's inherent uncertainty and large database, explorers "*get more unexpected results when creating images with it*" (PD2), which prompts them to "*step outside of bubble zones and ways of perceiving things*" (PD2). She illustrated this point by comparing AIGT and human creators involved in transforming place appearances: "Even with a highly creative partner, we may not create these alternative worlds because we're fixed in bubbles and same culture" (PD2). While AIGT encourages explorers to "think about familiar places differently" (PD2), it potentially causes misunderstandings about locations. PT2 highlighted this issue in her comment: "AI is sort of arranging stereotypes... It randomly interprets these scenes without telling the truth or guiding people to imagine better possibilities" (PT2). She gave the following example: "People may feel confused or think this is what Taiwan is like or what Holland is like. But in fact, this may not necessarily be the case" (PT2).

Interestingly, all the participants drew a comparison between city exploration with AIGT and Pokémon Go, a mainstream urban play allowing users to interact with the fictional Pokémon world in real-world locations. They remarked that Pokémon Go players mainly focus on screens instead of physical environments. As noted by them, the players mostly "search where spawns or gyms pop up" (PD1) and "go straight ahead to destinations without necessarily looking at the places around" (PD2). "Exploring the city is just an extra bonus in Pokémon Go" (PT1). Conversely, in this exploratory activity, the participants "spent more time paying attention to the surroundings" (PT1) and "photographing meaningful sceneries" (PD2) to obtain more intriguing results from AIGT. The process made them "look more at what's actually there and what you're actually capturing in the photo" (PD2).

# 4 Discussion and Future Work

## 4.1 Crafting Familiarization and Defamiliarization for Playful City Wandering with AIGT

Our research builds upon the initiatives of Playable Cities and urban play which use technologies to enable diverse urban experiences and city-making. We explored how AIGT drew upon urban appearances and the participants' place experiences to generate personalized and intriguing recontextualizations. From the findings, we identify two AIGT patterns: "familiarization," which transforms elements of places into those familiar to explorers, and "defamiliarization," which turns elements into those of unfamiliar worlds. These generated worlds can be either actual or fictional. AIGT retains resemblances in urban object shapes and photo compositions while manipulating the presence of non-existing elements to either align with or diverge from explorers' memories and impressions of places. This reinterpretation process could foster playfulness and provide opportunities for reimagining and rediscovering the surroundings, which opens up design space for meaning making and dialogue with cities. Yet, they also raise further research questions, including: How can (de)familiarization be used in urban play design forms, such as pervasive games and tangible installations? What strategies can be employed in the interaction involving both (un)familiar places? Who controls these strategies and to what extent? In future work, we plan to delve deeper into these questions through developing design interventions in field studies. Our goal is to propose a framework that assists designers in leveraging (de)familiarization to craft playful urban experiences empowered by AIGT.

## 4.2 AIGT as a 'Tourist' in Urban Play: Navigating Benefits and Risks

In the spectrum of (de)familiarizing places, we introduce the metaphor AIGT as a 'tourist' to discuss the design concern of applying AIGT to playful urban interaction. This term encapsulates AIGT's tendency to be unfamiliar with exact contexts depicted in input photos and prompts, attributed to limitations of the selected models and training data. Regarding design opportunities, the unfamiliarity of a 'tourist' renders the generation perceptible yet unpredictable, which motivates explorations in urban areas, observations of surprising substitutions, or speculations on the potentials of cityscapes. By connecting the generated results with their prompts and models, explorers could also (re)discover their relationships with the urban from an AI perspective. Nevertheless, AIGT as a 'tourist' also encounters unfamiliar territories and inadvertently stereotypes them. As articulated by the participants, AIGT's random juxtaposition of unfamiliar elements without sufficient explanation may lead to disappointing experiences or reinforce stereotypical impressions. We suggest that when applying AIGT to playable urban interactions, designers and researchers should scrutinize the backgrounds and consequences of this unfamiliarity. We also advocate for additional investigations and design interventions to mitigate misunderstandings and stereotypes resulting from AIGT content in urban settings.

# 5 Conclusion

In this exploratory study, we probe into the experience and design potentials of city exploration with AIGT. We present how participants (re)discovering the city when interacting with AIGT, the difference in their experience resulting from their familiarity with places, and the two roles of AIGT in city exploration. These findings provide initial empirical evidence of playfulness and reflection collectively formed by explorers, AIGT content, and urban space. We also contribute to design considerations, including (de)familiarizing places with AIGT and viewing it as a 'tourist,' which assist designers and researchers in leveraging and evaluating generative AI for crafting urban experiences.

## Acknowledgments

We extend our gratitude to all participants for generously contributing their time and for openly and actively providing feedback during the exploratory activity.